\documentclass[english,twocolumn,showpacs,preprintnumbers,amsmath,amssymb,aps,dblfloatfix]{revtex4-1}
\usepackage[T1]{fontenc}
\usepackage[latin9]{inputenc}
\usepackage{color}
\usepackage{array}
\usepackage{longtable}
\usepackage{units}
\usepackage{multirow}
\usepackage{amsmath}
\usepackage{amssymb}
\usepackage{graphicx}

\makeatletter

\providecommand{\tabularnewline}{\\}

\@ifundefined{textcolor}{}
{%
 \definecolor{BLACK}{gray}{0}
 \definecolor{WHITE}{gray}{1}
 \definecolor{RED}{rgb}{1,0,0}
 \definecolor{GREEN}{rgb}{0,1,0}
 \definecolor{BLUE}{rgb}{0,0,1}
 \definecolor{CYAN}{cmyk}{1,0,0,0}
 \definecolor{MAGENTA}{cmyk}{0,1,0,0}
 \definecolor{YELLOW}{cmyk}{0,0,1,0}
}

\usepackage{lmodern}
\usepackage[T1]{fontenc}
\usepackage{hyperref}
\usepackage{longtable}

\makeatother

\usepackage{babel}
\begin{document}

\title{ State of the art for \emph{ab initio} vs empirical potentials for
predicting 6e$^{-}$ excited state molecular energies: Application
to Li$_{2}\left(b,1^{3}\Pi_{u}\right)$}

\author{Nikesh S. Dattani,$^{1,2,3,\,}$}

\email{dattani.nike@gmail.com }

\affiliation{}

\affiliation{$^{1}$Quantum Chemistry Laboratory, Department of Chemistry, Kyoto
University, 606-8502, Kyoto, Japan,}

\affiliation{$^{2}$School of Materials Science and Engineering, Nanyang Technological
University, 639798, Singapore,}

\affiliation{$^{3}$Fukui Institute for Fundamental Chemistry, 606-8103, Kyoto,
Japan,}

\author{Robert J. Le~Roy,$^{3,\,}$}

\email{rleroy@uwaterloo.ca}

\affiliation{$^{4}$Guelph-Waterloo Centre for Graduate Work in Chemistry and
Biochemistry, University of Waterloo, N2L 3G1, Waterloo, Ontario,
Canada.}
\begin{abstract}
We build the first analytic empirical potential for the most deeply
bound Li$_{2}$ state: $b\left(1^{3}\Pi_{u}\right)$. Our potential
is based on experimental energy transitions covering $v=0-34$, and
very high precision theoretical long-range constants. It provides
high accuracy predictions up to $v=100$ which pave the way for high-precision
long-range measurements, and hopefully an eventual resolution of the
age old discrepancy between experiment and theory for the Li$\left(2^{2}S\right)+\mbox{Li}\left(2^{2}P\right)$
$C_{3}$ value. State of the art \emph{ab initio} calculations predict
vibrational energy spacings that are all in at most 0.8~cm$^{-1}$
disagreement with the empirical potential. 
\end{abstract}


\maketitle
There is currently a rather large discrepancy between the best atomic
3e$^{-}$\emph{ ab initio }calculation \cite{Tang2010}, and the most
current empirical value \cite{Gunton2013,LeRoy2009}, for the leading
long-range Li$(2^{2}S)-\mbox{Li\ensuremath{(2^{2}P)}}$ interaction
constant ($C_{3}$), despite the latter being the most precise experimentally
determined oscillator strength for any system, by an order of magnitude
\cite{Tang2011}. $\mbox{Li}_{2}(b,1^{3}\Pi_{u})$ is one of the molecular
states that dissociates to Li$(2^{2}S)+\mbox{Li\ensuremath{(2^{2}P)}}$,
and therefore its long-range potential has this $C_{3}$ interaction
constant. This ``$b$-state''\textbf{ }is also the deepest Li$_{2}$
potential, and out of the five lowest Li$_{2}$ states, the $b$-state
is the only one for which an analytic empirical potential has never
been made. 

Since the highest 2000 cm$^{-1}$ worth of vibrational levels of the
$b$-state still have not been observed, and part of this region is
now accessible by current ultra-high precision PA (photoassociation)
technology \cite{Gunton2013,Semczuk2013}, an analytic potential would
be very useful for making predictions to assist in observing the missing
levels. The $b$-state also mixes strongly with the $A(1^{1}\Sigma_{u})$-state
, which has by far the most precisely determined excited molecular
state potential in all of chemistry, yet still has a rather large
gap of missing data in the middle of its energy range \cite{Gunton2013,LeRoy2009}. 

Finally, the $b(1^{3}\Pi_{u})$-state has been a key doorway to the
triplet manifold, and was directly involved in the measurements for
a vast number of other triplet states such as $2^{3}\Delta_{g}$
\cite{Yi2001}, $3^{3}\Pi_{g}$ \cite{Yiannopoulou1995}, $3^{3}\Sigma_{g}^{+}$
\cite{Yiannopoulou1995,Li2007,Xie1986}, $2^{3}\Sigma_{g}^{+}$ \cite{Yiannopoulou1995,Lazarov2001},
$2^{3}\Pi_{g}$ \cite{Xie1986,Yiannopoulou1995,Russier1997,Li2007},
$1^{3}\Delta_{g}$ \cite{Xie1986,Linton1992,Yiannopoulou1995d,Weyh1996a,Russier1997,Li2007},
$1^{3}\Sigma_{g}^{-}$, and other undetermined $^{3}\Lambda$ states
\cite{Rai1985,Li1996a}. Some of these more highly excited triplet
states (namely $3^{3}\Sigma_{g}^{+},$ $2^{3}\Sigma_{g}^{+},$ $2^{3}\Pi_{g},$
and $1^{3}\Delta_{g}$) are so thoroughly covered by these spectroscopic
measurements, that global empirical potentials can be built for them
too. For this, an analytic potential for the $b$-state would be used
as a base.

In this work we will build analytic empirical potentials for the $b$-states
of all stable homonuclear isotopologues of Li$_{2}$. Previous work
has shown that analytic empirical potentials for the $c(1^{3}\Sigma_{g})$-state
were able to predict energies correctly to about 1~cm$^{-1}$, in
the middle of a gap of $>5000$~cm$^{-1}$ where data were unavailable
\cite{Dattani2011,Semczuk2013}, and this was much better agreement
than was obtained with the most sophisticated Li$_{2}$ \emph{ab initio}
calculations of the time \cite{Halls2001}. 

It was recently shown that the best ground-state rotationless \emph{ab
initio }potentials for the 5e$^{-}$ molecules BeH, BeD, and BeT,
were able to predict vibrational energy spacings to within 1~cm$^{-1}$
for all measured energy levels except one. The $b$-state of Li$_{2}$
might be expected to be more challenging \emph{ab initio }because
it (1) is an excited state, (2) has one more e$^{-}$, and (3) involves
many more vibrational energies. We will therefore compare our analytic
empirical potentials for the $b$-state of $^{6,6}$Li$_{2}$ and
$^{7,7}$Li$_{2}$ with the most state-of-the-art \emph{ab initio
}calculations, which were published recently in \cite{Musia2014}.

Table \ref{tab:Summary-of-experiments} summarizes all experiments
we could find which provided information on rovibrational levels of
the $b$-state . Unfortunately attempts to recover the data from \cite{Engelke1983,Rai1985,Xie1986,Schmidt1988,Weyh1996b}
were unsuccessful, but we were still able to include all data from
the other experiments in our study. Furthermore, it is noted that
the $b$-state was also involved in various other studies \cite{Yiannopoulou1995,Yiannopoulou1995e,Li1999,Yi2001,Li2007}
but these just made use of rovibrational levels that were already
determined in the studies listed in Table \ref{tab:Summary-of-experiments},
in order to access levels of other electronic states.

\begin{table*}
\protect\caption{Summary of experiments involving $\mbox{Li}$$_{2}(b,1^{3}\Pi_{u})$\label{tab:Summary-of-experiments}}

\rule[0.1ex]{1\textwidth}{0.5pt}

\begin{centering}
\begin{tabular*}{1\textwidth}{@{\extracolsep{\fill}}ccccccccc}
\noalign{\vskip2mm}
Isotopes & Year & Type & States Involved & Unc. (cm$^{-1}$) & $v$ & \# Data & Included in dataset & Source\tabularnewline[2mm]
\hline 
\hline 
\noalign{\vskip2mm}
$^{7,7}$Li$_{2}$ & 1985 & LIF & $A(1^{1}\Sigma_{u}^{+}),b(1^{3}\Pi_{u})$  & $a$ & $8-25$ & 100s &  & {\footnotesize{}Preuss \& Baumgartner \cite{Preuss1985}}\tabularnewline
 & 1985 & OODR & $(5d\delta^{3}\Delta_{g})\rightarrow b(1^{3}\Pi_{u})$  & 1 & $0$ & 100s & - & {\footnotesize{}Rai et al \cite{Rai1985}}\tabularnewline
 & 1992 & PFOODR & $A(1^{1}\Sigma_{u}^{+}),b(1^{3}\Pi_{u})$ & ? & $19$ & 3 & - & {\footnotesize{}Li et al. \cite{Li1992}}\tabularnewline
 & 1996 & cw PFOODR & $({}^{3}\Lambda_{g})\rightarrow b(1^{3}\Pi_{u})$ & ? & $0,1,4$ & ? & - & {\footnotesize{}Li et al.\cite{Li1996a}}\tabularnewline
 & 1996 & CIF & $(1^{3}\Delta_{g})\rightarrow b(1^{3}\Pi_{u})$ & ? & $0-11$ & ? & - & {\footnotesize{}Weyh et al.\cite{Weyh1996b}}\tabularnewline
 & 1997 & PFOODR & $(2^{3}\Pi_{g})\rightarrow b(1^{3}\Pi_{u})$  & 0.005 & $1-27$ & 178 & $\checkmark$ & {\footnotesize{}Russier et al \cite{Russier1997}}\tabularnewline
 & 1997 & PFOODR & $(1^{3}\Delta_{g})\rightarrow b(1^{3}\Pi_{u})$  & 0.005 & $1-25$ & 234 & $\checkmark$ & {\footnotesize{}Russier et al \cite{Russier1997}}\tabularnewline
 & 1997 & CIF & $(1^{3}\Delta_{g})\rightarrow b(1^{3}\Pi_{u})$  & 0.005 & $0-7$ & 314 & $\checkmark$ & {\footnotesize{}Russier et al \cite{Russier1997}}\tabularnewline
 & 2001 & cw PFOODR & $A(1^{1}\Sigma_{u}^{+}),b(1^{3}\Pi_{u})$  & ? & $15,22$ & 2 & - & {\footnotesize{}Lazarov \& Lyyra \cite{Lazarov2001}}\tabularnewline[2mm]
$^{7,6}$Li$_{2}$ & 1985 & LIF & $A(1^{1}\Sigma_{u}^{+}),b(1^{3}\Pi_{u})$  & $a$ & $8-20$ & 100s &  & {\footnotesize{}Preuss \& Baumgartner\cite{Preuss1985}}\tabularnewline[2mm]
$^{6,6}$Li$_{2}$ & 1983 & CIF & $(3^{3}\Delta_{g})\rightarrow b(1^{3}\Pi_{u})$ & $0.3$ & $0-10$ & 240 & - & {\footnotesize{}Engelke \& Hage }\tabularnewline
 & 1985 & LIF & $A(1^{1}\Sigma_{u}^{+}),b(1^{3}\Pi_{u})$  & $a$ & $8-18$ & 100s &  & {\footnotesize{}Preuss \& Baumgartner\cite{Preuss1985}}\tabularnewline
 & 1985 &  & $A(1^{1}\Sigma_{u}^{+}),b(1^{3}\Pi_{u})$  & ? & $9,15$ & 2 & - & {\footnotesize{}Xie \& Field \cite{Xie1985,Xie1985c}}\tabularnewline
 & 1986 & PFOODR & $(2^{3}\Pi_{g})\rightarrow b(1^{3}\Pi_{u})$ & $0.5$ & $0-17$ & \textasciitilde{}170 & {\footnotesize{}32 lines recovered} & {\footnotesize{}Xie \& Field \cite{Xie1986} }\tabularnewline
 & 1986 & PFOODR & $(1^{3}\Delta_{g})\rightarrow b(1^{3}\Pi_{u})$ & $0.02-0.13$ & $0-11$ & ? &  & {\footnotesize{}Rice, Xie \& Field \cite{Rice1986,RiceStevenF.1986}}\tabularnewline
 & 1988 & CIF & $(1^{3}\Delta_{g})\rightarrow b(1^{3}\Pi_{u})$ & $0.2$ & $0-11$ & \textasciitilde{}200 & - & {\footnotesize{}Schmidt }\emph{\footnotesize{}et al. \cite{Schmidt1988}}\tabularnewline
 & 1992 & CIF & $(1^{3}\Delta_{g})\leftrightarrow b(1^{3}\Pi_{u})$ & $0.003-0.07$ & $0-9$ & 599 & $\checkmark$ & {\footnotesize{}Linton }\emph{\footnotesize{}et al. }{\footnotesize{}\cite{Linton1992}}\tabularnewline[2mm]
\hline 
\noalign{\vskip2mm}
\multicolumn{3}{c}{\textbf{TOTAL}} &  & $0.003-1$ & $0-27$ & ? & 1357 & {\footnotesize{}\cite{Linton1992}\cite{Russier1997}}\tabularnewline[2mm]
\hline 
\end{tabular*}
\par\end{centering}

\rule[0.1ex]{1\textwidth}{0.5pt}

\raggedright{}{\scriptsize{}\vspace{0.5mm}
}{\footnotesize{}$^{a}$The measurements were on $v$-levels of the
$A$-state, and information about the $b$-state $v$-levels that
perturbed those $A$-state levels was inferred indirectly.}
\end{table*}

\section{Hamiltonian}

The rovibrational energy levels and wavefunctions for isotopologue
$\alpha$ with reduced mass $\mu_{\alpha}$ are treated as the eigenvalues
and eigenfunctions in the effective radial Schroedinger equation:

\begin{widetext}

\begin{equation}
\left(-\frac{\hbar}{2\mu_{\alpha}}\frac{{\rm d}^{2}}{{\rm d}r^{2}}+V_{{\rm \alpha}}(r)+\frac{\hbar N(N+1)}{2\mu_{\alpha}r^{2}}\left(1+g_{{\rm \alpha}}(r)\right)\right)\psi_{v,N}(r)=E_{v,N}(r)\psi_{v,N}\,.\label{eq:hamiltonian}
\end{equation}

\end{widetext}

\noindent Here $V_{\alpha}(r)$ and $g_{\alpha}(r)$ represent the
``adiabatic'' potential and the ``non-adiabatic'' rotational $g$-factor.
The adiabatic potential can be represented as a ``Born-Oppenheimer''
potential (which is mass-independent), plus a (mass-dependent) shift
due to the diagonal correction to the Born-Oppenheimer approximation:

\begin{equation}
V_{\alpha}(r)\equiv V_{{\rm BO}}(r)+\Delta V_{\alpha}(r).
\end{equation}
The $\Delta V_{\alpha}(r)$ correction can be approximated by the
expectation value of the nuclear kinetic energy operator in the molecular
electronic wavefunction basis $\left\langle T_{{\rm nuc,\alpha}}\right\rangle $
\cite{McAlexander1996}. For homonuclear diatomics it is given by
\cite{VanVleck1936,Bunker1968,McAlexander1996}:

\begin{eqnarray}
\Delta V_{\alpha}(r) & = & \left\langle T_{{\rm nuc},\alpha}\right\rangle +\Delta_{2}V_{\alpha}(r)\\
\left\langle T_{{\rm nuc},\alpha}\right\rangle  & \equiv & Q_{\alpha}(r)+P_{\alpha}(R)+S_{\alpha}(R)\\
Q_{\alpha}(r) & \equiv & -\frac{\hbar^{2}}{2\mu_{\alpha}}\left\langle \frac{\partial^{2}}{\partial r^{2}}\right\rangle _{\alpha}\\
P_{\alpha}(r) & \equiv & \frac{\hbar^{2}}{2\mu_{\alpha}}\left\langle L_{x}^{2}+L_{y}^{2}\right\rangle _{\alpha}
\end{eqnarray}

\begin{eqnarray}
S_{\alpha}(r) & \equiv & -\frac{\hbar^{2}}{8\mu_{\alpha}}\left(\left\langle \sum_{i}\nabla_{i}^{2}\right\rangle +\left\langle \sum_{i\ne j}\nabla_{i}\nabla_{j}\right\rangle \right)_{\alpha}\\
 & = & \frac{m_{e}}{4\mu_{\alpha}}\left\langle T_{e,\alpha}\right\rangle -\frac{\hbar^{2}}{8\mu_{\alpha}}\left\langle \sum_{i\ne j}\nabla_{i}\nabla_{j}\right\rangle _{\alpha}\\
 & = & -\frac{m_{e}}{4\mu_{\alpha}}\left(V_{{\rm BO}}(r)+r\frac{\partial}{\partial r}V_{{\rm BO}}(r)\right)\label{eq:firstTermOfS(r)}\\
 &  & \:-\frac{\hbar^{2}}{8\mu_{\alpha}}\left\langle \sum_{i\ne j}\nabla_{i}\nabla_{j}\right\rangle _{\alpha}\\
 & \equiv & -\frac{m_{e}}{4\mu_{\alpha}}\left(V_{{\rm BO}}(r)+r\frac{\partial}{\partial r}V_{{\rm BO}}(r)\right)+\Delta S_{\alpha}(r).
\end{eqnarray}
where $z$ represents the internuclear axis, $L_{x}$ and $L_{y}$
are then projections of the total electronic orbital angular momentum,
$i,j$ represent indices for individual electrons of the molecule,
and the first term of $S_{\alpha}(r)$ has been expressed in terms
of the average electronic kinetic energy $\langle T_{e,\alpha}\rangle$
and then re-expressed in terms of $V_{{\rm BO}}(r)$ using the virial
theorem \cite{Slater1933,McAlexander1996}. We can define a long-range
term $P_{{\rm LR,\alpha}}(r)$ by evaluating $P_{\alpha}(r)$ in the
long-range Heitler-London basis, where electron overlap is zero. $P_{\alpha}(r)$
is then expressed as $P_{{\rm LR,\alpha}}(r)$ plus a correction $\Delta P_{\alpha}(r)$
\cite{McAlexander1996}:

\begin{eqnarray}
P_{\alpha}(r) & \equiv & P_{{\rm LR},\alpha}(r)+\Delta P_{\alpha}(r)\\
P_{{\rm LR,\alpha}}(r) & = & \frac{\hbar^{2}}{2\mu_{\alpha}r^{2}}\sum_{k}^{2}l_{k}(l_{k}+1),
\end{eqnarray}
where $l_{k}$ represents the orbital angular momentum of the electrons
in constituent atom $k$ of the molecule. Herein we restrict our attention
to the $b$-state of Li$_{2}$ which dissociates into Li$(S)$ + Li$(P)$:

\begin{eqnarray}
P_{{\rm LR,\alpha}}(r) & = & \frac{\hbar^{2}}{2\mu_{\alpha}r^{2}}\left(l_{s}(l_{s}+1)+l_{p}(l_{p}+1)\right)\label{eq:P_LR}\\
 & = & \frac{2\hbar^{2}}{2\mu_{\alpha}r^{2}}\,,\,\mbox{since }\left(l_{s}=0,l_{p}=1\right).
\end{eqnarray}

While we know that in the long-range limit, $Q_{\alpha}(r)$ will
become a constant \cite{McAlexander1996}, $\Delta P_{\alpha}(r)$
will be zero, and $\Delta S_{\alpha}(r)$ will be small \cite{McAlexander1996},
no other information about these terms is known. Therefore, we may
re-write the diagonal Born-Oppenheimer correction (DBOC) in terms
of what we know, and then represent these parts that we don't know,
by model functions $\tilde{S_{k}}(r)$ for each constituent atom $k$
of the molecule:

\begin{widetext}

\begin{eqnarray}
\Delta V_{\alpha}(r) & = & \frac{\hbar^{2}}{\mu_{\alpha}r^{2}}-\frac{m_{e}}{4\mu_{\alpha}}\left(V_{{\rm BO}}(r)+r\frac{\partial}{\partial r}V_{{\rm BO}}(r)\right)+{\color{blue}{\color{red}\left(\Delta S_{\alpha}(r)+\Delta P_{\alpha}(r)+Q_{\alpha}(r)+\Delta_{2}V_{\alpha}(r)\right)}}\label{eq:DeltaV}\\
 & \equiv & \frac{\hbar^{2}}{\mu_{\alpha}r^{2}}-\frac{m_{e}}{4\mu_{\alpha}}\left(V_{{\rm BO}}(r)+r\frac{\partial}{\partial r}V_{{\rm BO}}(r)\right)+\sum_{k}\frac{m_{e}}{M_{k}}\tilde{S_{k}}(r),\label{eq:firstTimeStildeIsIntroduced}
\end{eqnarray}

\end{widetext}

\noindent \begin{flushleft}
where $m_{e}$ is the electron mass and $M_{k}$ is the mass of the
$k^{{\rm th}}$ constituent nucleus of the molecule. Note that until
now, the terms containing $\tilde{S_{k}}(r)$ represented the entirety
of Eq. \ref{eq:firstTimeStildeIsIntroduced}, so less of $\Delta V_{\alpha}(r)$
was described by theoretically known expressions, and more was described
by empirical fitting functions \cite{Watson2004}. 
\par\end{flushleft}

The only part of the Hamiltonian in Eq. \ref{eq:hamiltonian} that
is missing is now the non-adiabatic term $g_{\alpha}(r)$. This is
often represented by model functions $\tilde{R}_{k}(r)$ for each
atom:

\begin{equation}
g_{\alpha}(r)\equiv\sum_{k}\frac{m_{e}}{M_{k}}\tilde{R}_{k}(r)\,.\label{eq:rotational_g_factor}
\end{equation}

\begin{figure*}
\protect\caption{When considering the data for all isotopologues, levels up to $v=32$
have been observed. For $^{6,6}$Li$_{2}$, levels $v=33-40$ are
accessible by Kirk Madison's group in University of British Columbia
with the laboratory and method used in studies of the $A\left(1^{1}\Sigma_{u}^{+}\right)$
\cite{Gunton2013} and $c\left(1^{3}\Sigma_{g}^{+}\right)$ \cite{Semczuk2013}
states. The remainder of the levels have not been observed.\label{fig:b} }

\includegraphics[width=1\textwidth]{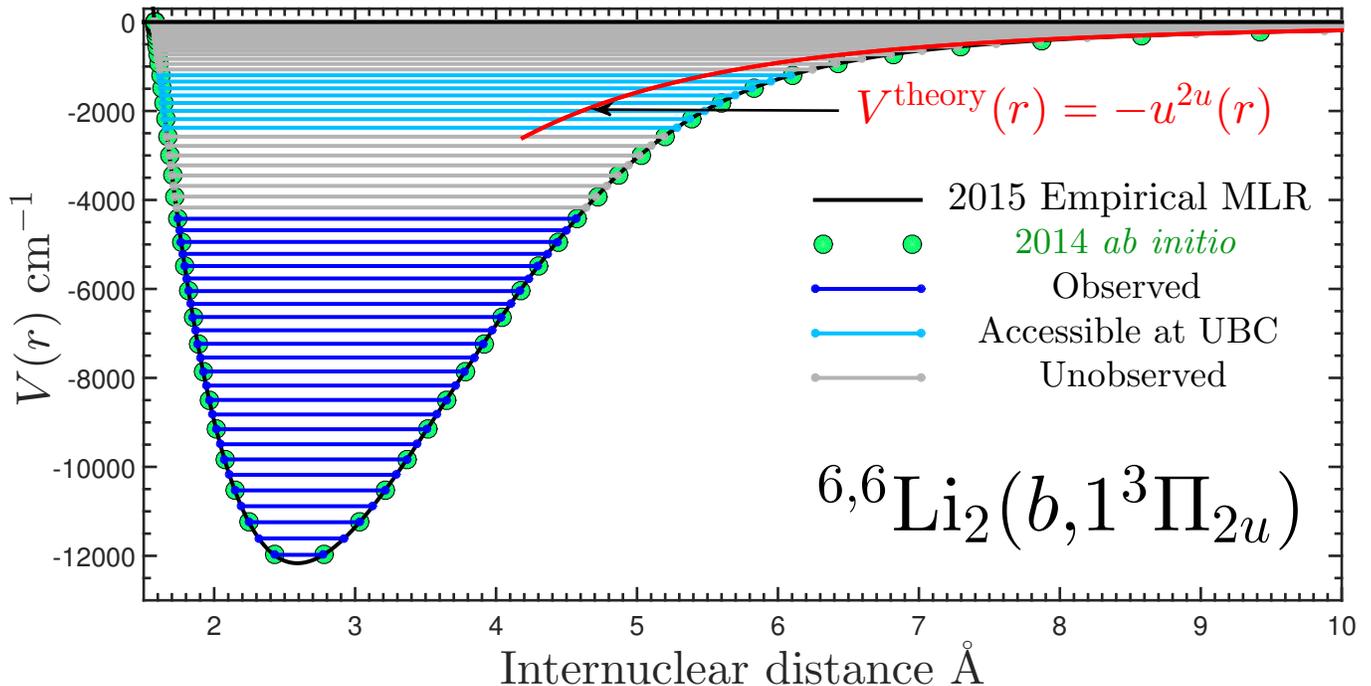}
\end{figure*}

\section{Empirical potential and Born-Oppenheimer breakdown (BOB) corrections}

We now wish to determine empirical functions for $V_{{\rm BO}}(r),\,\tilde{S}(r),$
and $\tilde{R}(r)$ that accurately reproduce all measured energies
when using the Hamiltonian of Eq. \ref{eq:hamiltonian}.

There is a gap of more than 2000$\:$cm$^{-1}$ ($>60$ THz) in experimental
information between the highest observed level of Li$_{2}\left(b,1^{3}\Pi_{u}\right)$,
and its dissociation energy. This means that when building an empirical
potential that aims to be relevant in the large data gap, it is very
important to take great care in ensuring the potential behaves physically
correctly in the extrapolation region. In 2011 the MLR (Morse/long-range)
model was used in a fit to build empirical potentials from spectroscopic
data for the $c\left(1^{3}\Sigma_{g}^{+}\right)-$states of $^{6,6}$Li$_{2}$
and $^{7,7}$Li$_{2}$, where there was a gap of more than $5000\:$cm$^{-1}$
between data near the bottom of the potential, and data at the very
top \cite{Dattani2011}. In 2013 spectroscopic measurements were made
in the very middle of this gap \cite{Semczuk2013}, and it was found
that the vibrational energies predicted by the MLR potential from
\cite{Dattani2011} were correct to within about 1$\:$cm$^{-1}$.
The present case for the $b$-state is in some sense more interesting
because there is no data at the top helping to anchor the potential
with the right shape near dissociation. 

However, as for the $c$-state, the MLR model is still expected to
be able to represent the physics in the extrapolation region faithfully
since the correct theoretical long-range is built into the model.
Having this long-range physics accurately built into the model is
almost as helpful as having data in the long-range region, as was
the case of the $c$-state. MLR-type empirical potentials have now
successfully described spectroscopic data for many diatomic \cite{LeRoy2006,LeRoy2007,Salami2007,Shayesteh2007,LeRoy2009,Coxon2010,Stein2010,Piticco2010a,LeRoy2011,Ivanova2011,Dattani2011,Xie2011,Yukiya2013,Knockel2013,Semczuk2013,Wang2013,Li2013,Gunton2013,Meshkov2014,Dattani2014c,Coxon2015,Walji2015,Dattani2015,Dattani2014b}
and polyatomic \cite{Li2008,Li2010,Tritzant-Martinez2013,Wang2013,Li2013a,Ma2014}
systems. Therefore, we will proceed to use the MLR model to describe
$V{\rm _{{\rm BO}}}(r)$.

The MLR model is defined by

\begin{equation}
V_{{\rm MLR}}(r)\equiv\mathfrak{D}_{e}\left(1-\frac{u(r)}{u(r_{e})}e^{-\beta(r)y_{p}^{r_{e}}(r)}\right)^{2},\label{eq:VMLR}
\end{equation}
where $\mathfrak{D}_{e}$ is the dissociation energy, $r_{e}$ is
the equilibrium internuclear distance, and the polynomial $\beta(r)$
is

{\scriptsize{}
\begin{equation}
\beta(r)\equiv\beta_{p,q}^{r_{{\rm ref}}}(r)\equiv\beta_{\infty}y_{p}^{r_{{\rm ref}}}(r)+\left(1-y_{p}^{r_{{\rm ref}}}(r)\right)\sum_{i=0}^{N_{\beta}}\beta_{i}\left(y_{q}^{r_{{\rm ref}}}(r)\right)^{i},\label{eq:betaPolynomial}
\end{equation}
}with 

\begin{equation}
\beta_{\infty}\equiv\lim_{r\rightarrow\infty}\beta(r)=\ln\left(\frac{2\mathfrak{D}_{e}}{u(r_{e})}\right).
\end{equation}
Equations \ref{eq:VMLR} and \ref{eq:betaPolynomial} also depend
on the radial variable

\begin{equation}
y_{n}^{r_{{\rm ref}}}(r)=\frac{r^{n}-r_{{\rm ref}}^{n}}{r^{n}+r_{{\rm ref}}^{n}},
\end{equation}
where the reference distance $r_{{\rm ref}}$ is simply the equilibrium
distance $r_{e}$ in most cases, but can be adjusted to optimize the
fit to equation \ref{eq:VMLR}.

It is well known \cite{LeRoy2009,Dattani2011} that for large $r$
we have 

\begin{equation}
V(r)\simeq\mathfrak{D}_{e}-u(r)+\cdots,\label{eq:VMLR_longRange}
\end{equation}
therefore the long-range behavior of the potential is defined by $u(r)$,
and the short to mid-range behavior is defined by $\beta(r).$ In
the $b\left(1^{3}\Pi_{u}\right)$ state, a spin-orbit interaction
emerges at large internuclear distances, which splits the potential
into four components. Therefore, four different potentials $V^{\Omega_{u}}(r)$
can be defined to have the same $\beta(r)$ defining the short-range
behavior where there is no significant splitting, and to have four
different $u^{^{3}\Pi_{u},\Omega_{u}}(r)$ defining the long-range
where the splitting occurs. 

For large $r$ where the spin-orbit interaction becomes strong, it
is dangerous to label the spin angular momentum $S$ and orbital angular
momentum $\Lambda$ separately, as in the molecular term symbol $^{2S+1}\Lambda$.
Instead, these two momenta are combined into a total electronic angular
momentum $\Omega$. For $^{3}\Pi_{u}$, $S=1$ and $\Lambda=1$, so
there are states with four possible symmetries in the $\Omega$ representation:
$0_{u}^{+}$, $0_{u}^{-}$, $1_{u}$, and $2_{u}$. Each of these
four states has a slightly different behavior at large internuclear
distances, due to coupling with states that have the same symmetry
in the $\Omega$ representation, but different symmetry in the $\Lambda$
representation. This coupling has been described in \cite{Aubert-Frecon1998}
and has been used for building appropriate analytic empirical potentials
for the $A\left(1^{1}\Sigma_{u}^{+}\right)$ \cite{LeRoy2009,Gunton2013}
and $c\left(1^{1}\Sigma_{u}^{+}\right)$ \cite{Dattani2011,Semczuk2013}
states of Li$_{2}$. The long-range function $u(r)$ is defined separately
for each spin-orbit state:

\begin{eqnarray}
u(r) & = & \begin{cases}
u^{b,0_{u}^{+}}(r) & ,\,0_{u}^{+}\\
u^{b,0_{u}^{-}}(r) & ,\,0_{u}^{-}\\
u^{b,1_{u}}(r) & ,\,1_{u}\\
u^{b,2_{u}}(r) & ,\,2_{u}.
\end{cases}
\end{eqnarray}
Each of these functions is an eigenvalue of a matrix $\mathbf{u}^{\Omega_{u}}$
for each $\Omega_{u}$ state. These matrices are given in the subsections
below, in terms of the (positive) spin-orbit splitting energy $\Delta E$,
and neglecting exchange interaction terms.

\subsection{The $0_{u}^{+}$ states}

In addition to the $b($1$^{3}\Pi_{u})$ state, the other $\Lambda$
state that can give rise to $0_{u}^{+}$ symmetry is the $A(1^{1}\Sigma_{u}^{+})$
state \cite{Jones2006}. The interstate coupling is therefore given
by the $2\times2$ matrix \cite{Aubert-Frecon1998}:

\vspace{5mm}

\begin{widetext}

\begin{eqnarray}
\mathbf{u}^{0_{u}^{+}}(r) & = & \begin{pmatrix}\frac{1}{3}{\displaystyle \sum_{\substack{m=3,6,8\\
9,10,11,\ldots
}
}}\frac{C_{m}^{A\left(1{}^{1}\Sigma_{u}^{+}\right)}+2C_{m}^{b\left(1{}^{3}\Pi_{u}\right)}}{r^{m}} & \quad\;\frac{\sqrt{2}}{3}{\displaystyle \sum_{\substack{m=3,6,8\\
9,10,11,\ldots
}
}}\frac{C_{m}^{A\left(1{}^{1}\Sigma_{u}^{+}\right)}-C_{m}^{b\left(1{}^{3}\Pi_{u}\right)}}{r^{3}}\\
\frac{\sqrt{2}}{3}{\displaystyle \sum_{\substack{m=3,6,8\\
9,10,11,\ldots
}
}}\frac{C_{m}^{A\left(1{}^{1}\Sigma_{u}^{+}\right)}-C_{m}^{b\left(1{}^{3}\Pi_{u}\right)}}{r^{m}} & \qquad\quad\Delta E+\frac{2}{3}{\displaystyle \sum_{\substack{m=3,6,8\\
9,10,11,\ldots
}
}}\frac{C_{m}^{A\left(1{}^{1}\Sigma_{u}^{+}\right)}+C_{m}^{b\left(1{}^{3}\Pi_{u}\right)}}{r^{m}}
\end{pmatrix},\label{eq:0+}
\end{eqnarray}

\noindent where the lower energy eigenvalue $u^{A,0_{u}^{+}}(r)$
comes from $A\left(1^{1}\Sigma_{u}^{+}\right)$ and approaches the
dissociation limit of Li$\left(2^{2}S_{\nicefrac{1}{2}}\right)+\mbox{Li}\left(2^{2}P{}_{\nicefrac{1}{2}}\right)$,
and the higher energy eigenvalue $u^{b,0_{u}^{+}}(r)$ comes from
$b\left(1^{3}\Pi_{u}\right)$ and approaches the dissociation limit
of Li$\left(2^{2}S_{\nicefrac{1}{2}}\right)+\mbox{Li}\left(2^{2}P{}_{\nicefrac{3}{2}}\right)$
\cite{Jones2006}.

\subsection{The $0_{u}^{-}$ states}

In addition to the $b($1$^{3}\Pi_{u})$ state, the other $\Lambda$
state that can give rise to $0_{u}^{-}$ symmetry is the $2a(2^{3}\Sigma_{u}^{+})$
state \cite{Jones2006}. The interstate coupling is therefore given
by the $2\times2$ matrix \cite{Aubert-Frecon1998}:

\begin{eqnarray}
\mathbf{u}^{0_{u}^{-}}(r) & = & \begin{pmatrix}\frac{1}{3}{\displaystyle \sum_{\substack{m=3,6,8\\
9,10,11,\ldots
}
}}\frac{C_{m}^{2a\left(2{}^{3}\Sigma_{u}^{+}\right)}+2C_{m}^{b\left(1{}^{3}\Pi_{u}\right)}}{r^{m}} & \quad\;\frac{\sqrt{2}}{3}{\displaystyle \sum_{\substack{m=3,6,8\\
9,10,11,\ldots
}
}}\frac{C_{m}^{2a\left(2{}^{3}\Sigma_{u}^{+}\right)}-C_{m}^{b\left(1{}^{3}\Pi_{u}\right)}}{r^{3}}\\
\frac{\sqrt{2}}{3}{\displaystyle \sum_{\substack{m=3,6,8\\
9,10,11,\ldots
}
}}\frac{C_{m}^{2a\left(2{}^{3}\Sigma_{u}^{+}\right)}-C_{m}^{b\left(1{}^{3}\Pi_{u}\right)}}{r^{m}} & \qquad\quad-\Delta E+\frac{2}{3}{\displaystyle \sum_{\substack{m=3,6,8\\
9,10,11,\ldots
}
}}\frac{C_{m}^{2a\left(2{}^{3}\Sigma_{u}^{+}\right)}+C_{m}^{b\left(1{}^{3}\Pi_{u}\right)}}{r^{m}}
\end{pmatrix},\label{eq:0-}
\end{eqnarray}

\noindent where the lower energy eigenvalue comes from $b\left(1^{3}\Pi_{u}\right)$
and approaches the dissociation limit of Li$\left(2^{2}S_{\nicefrac{1}{2}}\right)+\mbox{Li}\left(2^{2}P{}_{\nicefrac{1}{2}}\right)$,
and the higher energy eigenvalue comes from $2a\left(2^{3}\Sigma_{u}^{+}\right)$
and approaches the dissociation limit of Li$\left(2^{2}S_{\nicefrac{1}{2}}\right)+\mbox{Li}\left(2^{2}P{}_{\nicefrac{3}{2}}\right)$
\cite{Jones2006}.

\subsection{The $1_{u}$ states}

In addition to the $b($1$^{3}\Pi_{u})$ state, the other $\Lambda$
states that can give rise to $1_{u}$ symmetry are the $2a\left(2^{3}\Sigma_{u}^{+}\right)$
state and the $B\left(1{}^{1}\Pi_{u}\right)$state \cite{Jones2006}.
The interstate coupling is therefore given by the $3\times3$ matrix
\cite{Aubert-Frecon1998}:

\vspace{5mm}
\begin{equation}
\mathbf{u}^{1_{u}}(r)=
\end{equation}

{\tiny{}
\begin{equation}
\negthickspace\negthickspace\hspace{-3mm}\left(\begin{array}{ccc}
\frac{1}{3}{\displaystyle \sum_{m}\frac{C_{m}^{2a\left(2{}^{3}\Sigma_{u}^{+}\right)}+C_{m}^{B\left(1{}^{1}\Pi_{u}\right)}+C_{m}^{b\left(1{}^{3}\Pi_{u}\right)}}{r^{m}}} & \frac{1}{3\sqrt{2}}{\displaystyle \sum_{m}\frac{-2C_{m}^{2a\left(2{}^{3}\Sigma_{u}^{+}\right)}+C_{m}^{B\left(1{}^{1}\Pi_{u}\right)}+C_{m}^{b\left(1,^{3}\Pi_{u}\right)}}{r^{m}}} & \frac{1}{\sqrt{6}}{\displaystyle \sum_{m}\frac{-C_{m}^{B\left(1,^{1}\Pi_{u}\right)}+C_{m}^{b\left(1,^{3}\Pi_{u}\right)}}{r^{m}}}\\
\frac{1}{3\sqrt{2}}{\displaystyle \sum_{m}\frac{-2C_{m}^{2a\left(2{}^{3}\Sigma_{u}^{+}\right)}+C_{m}^{B\left(1{}^{1}\Pi_{u}\right)}+C_{m}^{b\left(1{}^{3}\Pi_{u}\right)}}{r^{m}}} & \Delta E_{{\rm SO}}+\frac{1}{6}{\displaystyle \sum_{m}\frac{4C_{m}^{2a\left(2{}^{3}\Sigma_{u}^{+}\right)}+C_{m}^{B\left(1{}^{1}\Pi_{u}\right)}+C_{m}^{b\left(1{}^{3}\Pi_{u}\right)}}{r^{m}}} & \frac{1}{2\sqrt{3}}{\displaystyle \sum_{m}\frac{-C_{m}^{B\left(1{}^{1}\Pi_{u}\right)}+C_{m}^{b\left(1{}^{3}\Pi_{u}\right)}}{r^{m}}}\\
\frac{1}{\sqrt{6}}{\displaystyle \sum_{m}\frac{-C_{m}^{B\left(1{}^{1}\Pi_{u}\right)}+C_{m}^{b\left(1{}^{3}\Pi_{u}\right)}}{r^{m}}} & \frac{1}{2\sqrt{3}}{\displaystyle \sum_{m}\frac{-C_{m}^{B\left(1{}^{1}\Pi_{u}\right)}+C_{m}^{b\left(1{}^{3}\Pi_{u}\right)}}{r^{m}}} & -\Delta E+\frac{1}{2}{\displaystyle \sum_{m}\frac{C_{m}^{B\left(1{}^{1}\Pi_{u}\right)}+C_{m}^{b\left(1{}^{3}\Pi_{u}\right)}}{r^{m}}}
\end{array}\right),\label{eq:1u}
\end{equation}
}{\tiny \par}

\noindent where the lowest energy eigenvalue comes from $b\left(1^{3}\Pi_{u}\right)$
and approaches the dissociation limit of Li$\left(2^{2}S_{\nicefrac{1}{2}}\right)+\mbox{Li}\left(2^{2}P{}_{\nicefrac{1}{2}}\right)$,
and the middle and highest energy eigenvalues come from $B\left(1^{1}\Pi_{u}\right)$
and $2a\left(2^{3}\Sigma_{u}^{+}\right)$ respectively and both approach
the dissociation limit of Li$\left(2^{2}S_{\nicefrac{1}{2}}\right)+\mbox{Li}\left(2^{2}P{}_{\nicefrac{3}{2}}\right)$
\cite{Jones2006}.

\subsection{The $2_{u}$ state}

The $2_{u}$ state approaching the dissociation limit of $2S+2P$
is alone in its symmetry, and approaches Li$\left(2^{2}S_{\nicefrac{1}{2}}\right)+\mbox{Li}\left(2^{2}P{}_{\nicefrac{3}{2}}\right)$
\cite{Jones2006}. It has the long-range function \cite{Tang2011}:

\begin{eqnarray}
u^{2_{u}}(r) & = & -\left(\Delta E-\sum_{\substack{m=3,6,8,\\
9,10,11,\ldots
}
}\frac{C_{m}^{b\left(1{}^{3}\Pi_{u}\right)}}{r^{m}}\right)\label{eq:2u}\\
 & = & -\left(\Delta E-\frac{C_{3}^{b\left(1{}^{3}\Pi_{u}\right)}}{r^{3}}-\frac{C_{6}^{b\left(1{}^{3}\Pi_{u}\right)}}{r^{6}}-\frac{C_{8}^{b\left(1{}^{3}\Pi_{u}\right)}}{r^{8}}-\frac{C_{9}^{b\left(1{}^{3}\Pi_{u}\right)}}{r^{9}}-\frac{C_{10}^{b\left(1{}^{3}\Pi_{u}\right)}}{r^{10}}-\frac{C_{11}^{b\left(1{}^{3}\Pi_{u}\right)}}{r^{11}}\cdots.\right)
\end{eqnarray}

\end{widetext}

\subsection{All four $\Omega_{u}$ states combined}

One can imagine an experiment which obtains spectroscopic measurements
for all of the four $\Omega_{u}$ states, and fits to all of this
data simultaneously by using the appropriate eigenvalues of the $8\times8$
matrix below:\vspace{-5mm}

\begin{eqnarray}
u & = & \left(\begin{array}{cccc}
\mathbf{u^{0_{u}^{+}}}\\
 & \mathbf{u^{0_{u}^{-}}}\\
 &  & \mathbf{u^{1_{u}}}\\
 &  &  & u^{2_{u}}
\end{array}\right)\\
 & = & \left(\begin{array}{cccccccc}
u_{11}^{0_{u}^{+}} & u_{12}^{0_{u}^{+}}\\
u_{21}^{0_{u}^{+}} & u_{22}^{0_{u}^{+}}\\
 &  & u_{11}^{0_{u}^{-}} & u_{12}^{0_{u}^{-}}\\
 &  & u_{21}^{0_{u}^{-}} & u_{22}^{0_{u}^{-}}\\
 &  &  &  & u_{11}^{1_{u}} & u_{12}^{1_{u}} & u_{13}^{1_{u}}\\
 &  &  &  & u_{21}^{1_{u}} & u_{22}^{1_{u}} & u_{23}^{1_{u}}\\
 &  &  &  & u_{31}^{1_{u}} & u_{32}^{1_{u}} & u_{33}^{1_{u}}\\
 &  &  &  &  &  &  & u^{2_{u}}
\end{array}\right).\label{eq:fullInteractionMatrix}
\end{eqnarray}

\begin{figure*}
\protect\caption{In the long-range region, the splitting of the $b\left(1^{3}\Pi_{u}\right)$
state into four spin-orbit components becomes increasingly obvious.
This is well beyond the region where data is available, so our empirical
potential fits to a model with the simplest long-range potential energy
function (the $2_{u}$ state, since there are no other $2_{u}$ states
that are nearby in energy, see Eqs. \ref{eq:2u}-\ref{eq:fullInteractionMatrix}).
The fine-structure splitting of $0.33532461313$~cm$^{-1}$ comes
from measurements in \cite{Brown2013}. \label{fig:long-range}}

\includegraphics[width=1\textwidth]{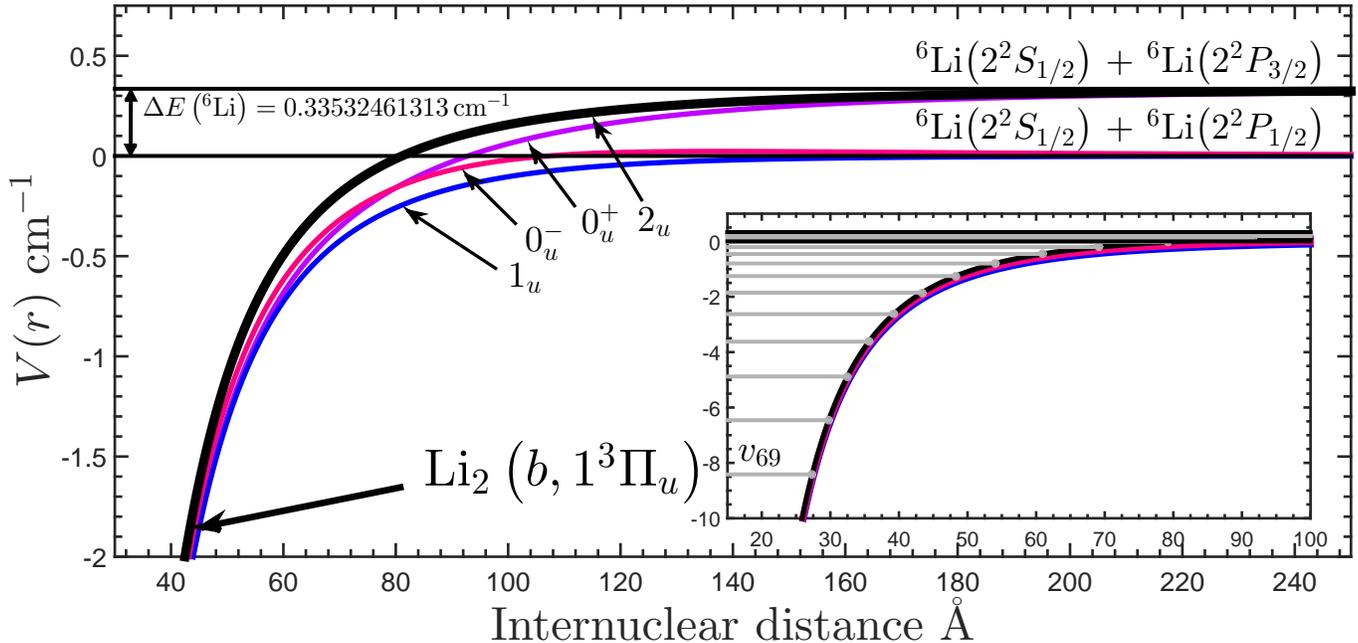}
\end{figure*}

However, Fig. \ref{fig:b} shows the data region, and Fig. \ref{fig:long-range}
shows that the spin-orbit splitting does not seem to become apparent
until well past this region. Since the measurements that have been
done on the $b$-state thus far are far away from the effect of the
spin-orbit splitting, we choose to use the simplest spin-orbit long-range
function: $u^{b,2_{u}}(r)$.

\subsection{Quadratic corrections and damping functions}

Since the leading term not shown in Eq. \ref{eq:VMLR_longRange} is
$\frac{u(r)^{2}}{4\mathfrak{D}_{e}}$, the contribution of the $C_{3}$
terms to the long-range form of the potential, will interfere with
the desired $C_{6}$ and $C_{8}$ terms, and all $C_{9}$ and $C_{11}$
terms will therefore have spurious contributions from the cross-terms
formed by the products of the $C_{3}$ terms with the $C_{6}$ and
$C_{8}$ terms respectively. We fix this in the same way as was done
for $C_{6}$ and $C_{9}$ in \cite{Dattani2008,LeRoy2009,Dattani2011,Semczuk2013,Gunton2013},
 by applying a transformation to all $C_{6}$, $C_{9}$, and this
time also $C_{11}$ terms:

\begin{eqnarray}
C_{6} & \rightarrow & C_{6}+\frac{C_{3}^{2}}{4\mathfrak{D}_{e}}\label{eq:C6dattaniCorrction}\\
C_{9} & \rightarrow & C_{9}+\frac{C_{3}C_{6}}{2\mathfrak{D}_{e}},\label{eq:C9dattaniCorrection}\\
C_{11} & \rightarrow & C_{11}+\frac{C_{3}C_{8}}{2\mathfrak{D}_{e}}.
\end{eqnarray}

{\scriptsize{}}{\scriptsize \par}

{\scriptsize{}}where the transformation in Eq. \ref{eq:C6dattaniCorrction}
has to be made first due to Eq. \ref{eq:C9dattaniCorrection}'s dependence
on $C_{6}$.

Additionally, The long-range formulas in terms of $C_{m}$ constants
in the above sub-sections were derived under the assumption that two
free atoms are interacting with each other, and there is no overlap
of the electrons' wavefunctions as in a bound molecule. To take into
account the effect of electron overlap, we use the damping function
form from \cite{LeRoy2011}:

\begin{eqnarray}
C_{m} & \rightarrow & C_{m}D_{m}^{(s)}\left(r\right)\\
D_{m}^{(s)}\left(r\right) & \equiv & \left(1-e^{-\left(\frac{b^{(s)}\rho r}{m}+\frac{c^{(s)}\left(\rho r\right)^{2}}{\sqrt{m}}\right)}\right)^{m+s},\label{eq:damping (last MLR definition)}
\end{eqnarray}

\noindent where for interacting atoms A and B, $\rho\equiv\rho_{{\rm AB}}=\frac{2\rho_{{\rm A}}\rho_{{\rm B}}}{\rho_{{\rm A}}\rho_{{\rm B}}},$
in which $\rho_{{\rm X}}\equiv\left(\nicefrac{I^{{\rm X}}}{I^{{\rm H}}}\right)^{\nicefrac{2}{3}}$
is defined in terms of the ionization potentials of atom X, denoted
$\left(I^{{\rm X}}\right)$, and hydrogen$\left(I^{{\rm H}}\right)$.
We use $s=-1$, which as shown in \cite{LeRoy2011}, means that the
MLR potential in Eq. \ref{eq:VMLR} has the physically desired behavior
$V\propto\nicefrac{1}{r^{2}}$ in the limit as $r\rightarrow0$. For
$s=-1$, the system independent parameters take the values $b^{(-1)}=3.30$,
and $c^{(-1)}=0.423$ \cite{LeRoy2011}.

\subsection{Long-range constants}

In previous studies of the $A(1^{1}\Sigma_{u}^{+})$ state \cite{LeRoy2009,Gunton2013b}
and $c(1^{3}\Sigma_{g}^{+})$ state \cite{Dattani2011,Semczuk2013},
it was found that the most precise theoretical values of $C_{3}$
known at those times \cite{Tang2010a,Tang2009b} did not fit as well
with the measurements of the high-lying vibrational levels near the
dissociation, as the values of $C_{3}$ obtained by setting it as
a free parameter determined by a least-squares fit to the data.  

However, for the $b\left(1^{3}\Pi_{u}\right)$-state, no measurements
of such high-lying vibrational levels have been made, so such an ``empirical
fit'' to $C_{3}$ is impossible, and we will have to use the most
precise theoretical value known. For $^{7,7}$Li$_{2}$ this is the
value from \cite{Tang2010a} and for $^{6,6}$Li$_{2}$ this is an
unpublished value from Tang calculated in 2015 \cite{Tang2015}. These
values are listed in Table \ref{tab:longRangeConstants}, along with
the theoretical values for the higher-order $C_{m}$ constants used
in our analysis (it has not yet been possible to fit these higher-order
$m>3$ constants to spectroscopic data in any direct-potential-fit
analysis, so they are held fixed). For $m>8$, no finite-mass corrections
have been calculated yet. 

\begin{table*}
\protect\caption{The best currently available long-range constants and their sources
(in Hartree atomic units). $^{\infty}$Li$_{2}$ represents a Li$_{2}$
molecule where both nuclei have infinite mass, since finite mass corrections
have not yet been calculated for $C_{m}$ coefficients with $m\ge9$.
\label{tab:longRangeConstants}}

\begin{tabular*}{1\textwidth}{@{\extracolsep{\fill}}cccccccccc}
\hline 
\noalign{\vskip2mm}
\multirow{1}{*}{} & \multicolumn{2}{c}{$A\left(1^{1}\Sigma_{u}\right)$} & \multicolumn{2}{c}{$2a\left(2^{3}\Sigma_{u}\right)$} & \multicolumn{2}{c}{$B\left(1^{1}\Pi_{u}\right)$} & \multicolumn{2}{c}{$b\left(1^{3}\Pi_{u}\right)$} & \tabularnewline[2mm]
 & $^{6,6}$Li$_{2}$ & $^{7,7}$Li$_{2}$ & $^{6,6}$Li$_{2}$ & $^{7,7}$Li$_{2}$ & $^{6,6}$Li$_{2}$ & $^{7,7}$Li$_{2}$ & $^{6,6}$Li$_{2}$ & $^{7,7}$Li$_{2}$ & {\small{}Ref.}\tabularnewline[2mm]
\hline 
\noalign{\vskip2mm}
{\footnotesize{}$C_{3}$} & {\footnotesize{}$11.0009$ \cite{Tang2015}} & {\footnotesize{}$11.0007$ \cite{Tang2015}} & {\footnotesize{}$-11.0009$ \cite{Tang2015}} & {\footnotesize{}$-11.0007$ \cite{Tang2010a}} & {\footnotesize{}~~$-5.5005$ \cite{Tang2015}} & {\footnotesize{}~~$-5.5004$ \cite{Tang2015}} & {\footnotesize{}~~5.5005 \cite{Tang2015}} & {\footnotesize{}~~5.5004 \cite{Tang2015}} & -\tabularnewline
{\footnotesize{}$C_{6}$} & {\footnotesize{}2076.19(7)} & {\footnotesize{}2076.08(7)} & {\footnotesize{}2076.19(7)} & {\footnotesize{}2076.08(7)} & {\footnotesize{}1407.20(2)} & {\footnotesize{}1407.15(5)} & {\footnotesize{}1407.20(2)} & {\footnotesize{}1407.15(5)} & {\footnotesize{}\cite{Tang2009}}\tabularnewline
{\footnotesize{}$C_{8}$} & {\footnotesize{}274137(6)} & {\footnotesize{}274128(5)} & {\footnotesize{}991104(5)} & {\footnotesize{}991075(6)} & {\footnotesize{}48566.9(4)} & {\footnotesize{}48566.4(2)} & {\footnotesize{}103053(1)} & {\footnotesize{}103052(1)} & {\footnotesize{}\cite{Tang2009}}\tabularnewline[2mm]
 & \multicolumn{2}{c}{{\footnotesize{}$^{\infty}$Li$_{2}$}} & \multicolumn{2}{c}{{\footnotesize{}$^{\infty}$Li$_{2}$}} & \multicolumn{2}{c}{{\footnotesize{}$^{\infty}$Li$_{2}$}} & \multicolumn{2}{c}{{\footnotesize{}$^{\infty}$Li$_{2}$}} & \tabularnewline[2mm]
{\footnotesize{}$C_{9}$} & \multicolumn{2}{c}{{\footnotesize{}2.2880(2)$\times10^{5}$}} & \multicolumn{2}{c}{{\footnotesize{}$-2.2880(2)$$\times10^{5}$}} & \multicolumn{2}{c}{{\footnotesize{}$-5.173(1)$$\times10^{4}$}} & \multicolumn{2}{c}{{\footnotesize{}5.173(1)$\times10^{4}$}} & {\footnotesize{}\cite{Tang2011}}\tabularnewline
{\footnotesize{}$C_{10}$} & \multicolumn{2}{c}{{\footnotesize{}3.0096$\times10^{7}$}} & \multicolumn{2}{c}{{\footnotesize{}$-1.2113$$\times10^{8}$}} & \multicolumn{2}{c}{{\footnotesize{}8.9295$\times10^{6}$}} & \multicolumn{2}{c}{{\footnotesize{}9.1839$\times10^{5}$}} & {\footnotesize{}\cite{Zhang2007}}\tabularnewline
{\footnotesize{}$C_{11}$} & \multicolumn{2}{c}{{\footnotesize{}$-5.930\times10^{7}$}} & \multicolumn{2}{c}{{\footnotesize{}$-5.0321$$\times10^{8}$}} & \multicolumn{2}{c}{{\footnotesize{}$-9.924\times10^{7}$}} & \multicolumn{2}{c}{{\footnotesize{}2.652$\times10^{7}$}} & {\footnotesize{}\cite{Tang2011}}\tabularnewline[2mm]
\hline 
\end{tabular*}

\rule[0.1ex]{1\textwidth}{0.5pt}
\end{table*}

\begin{figure*}
\protect\caption{Various $(N_{\beta})_{p,q}^{r_{{\rm ref}}}$ models with the same
quality of fit are in agreement with each other, until large values
of $r$ at which the $N_{\beta}=7$ model dips significantly below
the red curve representing the theoretical long-range behavior. This
dip was obvious in the $-r^{3}V(r)$ vs $\nicefrac{1}{r^{3}}$ plots
for all models mentioned in the text except for the three $N_{\beta}=11$
cases shown. The inset shows $-r^{3}V(r)$ for the four mentioned
models, subtracted from the red theoretical curve. All models match
the theory for very large $r$ (small $\nicefrac{1}{r^{3}}$), but
the cases with higher $p$ and/or $q$ match the theory for a much
larger range of $r$. Each green circle represents an outer classical
turning point of the \emph{ab initio }potential. In the inset, the
\emph{ab initio }values lie too high above 0 to be seen. \label{fig:leroySpace}}

\includegraphics[width=1\textwidth]{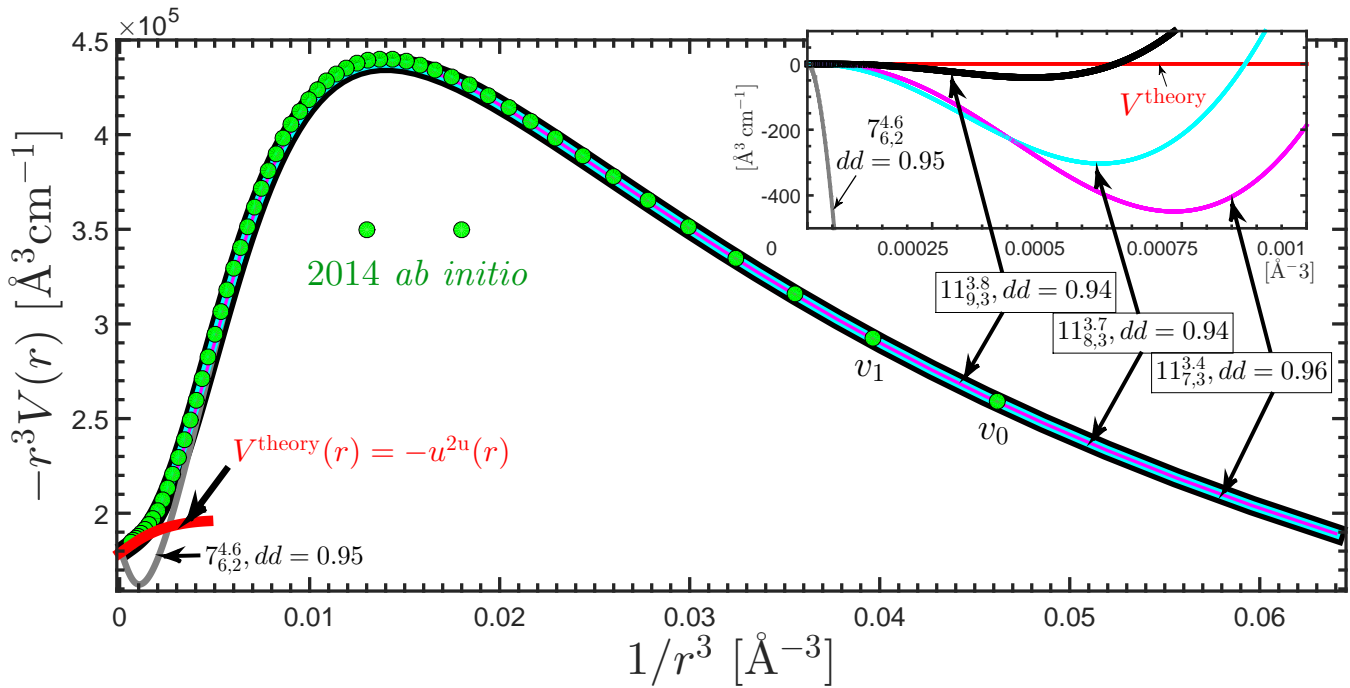}
\end{figure*}

\subsection{Dissociation energy $\mathfrak{D}_{e}$}

At the time of carrying out our analysis, the best experimental value
for $\mathfrak{D}_{e}$ of which we were aware, was the 1983 value
from \cite{Engelke1983}: 12145$\pm$200~cm$^{-1}$. In a recent
study on BeH \cite{Dattani2015}, the gap between the highest observed
level and the dissociation asymptote was $\sim1000$~cm$^{-1}$,
and the fitted value of $\mathfrak{D}_{e}$ varied by about $400$~cm$^{-1}$
as parameters such as $r_{{\rm ref}}$, $p$, and $q$ were changed.
For the present case of the $b$-state of Li$_{2}$, the data region
stops more than 4000~cm$^{-1}$ below the dissociation asymptote,
so we do not expect to be able to determine $\mathfrak{\mathfrak{D}_{e}}$any
more precisely than the 1983 experimental value. However, we still
tried, by letting $\mathfrak{D}_{e}$ be a free parameter, and we
indeed found that the fitted values varied by more than $400$~cm$^{-1}$.
Therefore, it might make sense to use experimental value from \cite{Engelke1983}
which was claimed  to be within 400~cm$^{-1}$.

However, it is expected that the \emph{ab initio} value from \cite{Musia2014}
correct to within much less than 400~cm$^{-1}$. This is because
we systematically checked the \emph{ab initio }$\mathfrak{D}_{e}$
values for all electronic Li$_{2}$ states calculated in \cite{Musia2014},
and found that they were at most 68~cm$^{-1}$ different from the
best experimental value, even when the experimental values were known
to as high of a precision as 0.0023~cm$^{-1}$ (see Table \ref{tab:Dissociation-energies}).
Furthermore, the \emph{ab initio }value for the $b$-state was within
the 400~cm$^{-1}$ confidence interval given by the 1983 experimental
value \cite{Engelke1983} discussed in the previous paragraph. Therefore,
we decided to fix our $\mathfrak{\mathfrak{D}_{e}}$ value at the
\emph{ab initio} value of 12166~cm$^{-1}$ and to only allow the
other parameters be free parameters for the remainder of the fitting
analysis. 

After the completion of this work, we discovered that a much less
known paper co-authored by one of the same authors from \cite{Engelke1983},
reported a more precise $\mathfrak{D}_{e}$ value of (12180.6$\pm0.6$)~cm$^{-1}$
just over 4 years afterwards \cite{Schmidt1988}, but it is not clear
in the paper how this value, nor its uncertainty is obtained. Particularly,
it is not clear whether this is a purely empirical value, or if it
also uses the \emph{ab inito} potential which is part of the subject
of the paper.

\begin{table*}
\protect\caption{Dissociation energies $\mathfrak{D}_{e}$ in cm$^{-1}$ of various
electronic states of Li$_{2}$. ``obs - calc'' denotes the \emph{ab
initio }value subtracted from the empirical value. Numbers in parenthesis
indicate the experimental uncertainty in the last digit(s) shown.
Where possible, $^{6,6}$Li$_{2}$ values were used. for the empirical
values. When empirical values for the $\mathfrak{D}_{e}$ of $^{6,6}$Li$_{2}$
were not available, the empirical value for $^{7,7}$Li$_{2}$ is
given. All \emph{ab initio} values are for $^{\infty}$Li$_{2}$ which
is a hypothetical Li$_{2}$ molecule with infinite mass. \label{tab:Dissociation-energies}}

\begin{tabular*}{1\textwidth}{@{\extracolsep{\fill}}cr@{\extracolsep{0pt}.}lr@{\extracolsep{0pt}.}lr@{\extracolsep{0pt}.}lr@{\extracolsep{0pt}.}lr@{\extracolsep{0pt}.}lr@{\extracolsep{0pt}.}l}
\hline 
\noalign{\vskip2mm}
\textbf{\textcolor{blue}{$2S+2S$}} & \multicolumn{2}{c}{$X(1^{1}\Sigma_{g}^{+})$} & \multicolumn{2}{c}{$a(1^{3}\Sigma_{u}^{+})$} & \multicolumn{2}{c}{} & \multicolumn{2}{c}{} & \multicolumn{2}{c}{} & \multicolumn{2}{c}{}\tabularnewline[2mm]
\hline 
\noalign{\vskip2mm}
\emph{ab initio }\cite{Musia2014} & 8466& & 334& & \multicolumn{2}{c}{} & \multicolumn{2}{c}{} & \multicolumn{2}{c}{} & \multicolumn{2}{c}{}\tabularnewline
empirical & 8516&7800(23) \cite{Gunton2013} & 333&7795(62) \cite{Semczuk2013} & \multicolumn{2}{c}{} & \multicolumn{2}{c}{} & \multicolumn{2}{c}{} & \multicolumn{2}{c}{}\tabularnewline
obs - calc & 51& & 0& & \multicolumn{2}{c}{} & \multicolumn{2}{c}{} & \multicolumn{2}{c}{} & \multicolumn{2}{c}{}\tabularnewline[2mm]
\hline 
\noalign{\vskip2mm}
\textbf{\textcolor{blue}{$2S+2P$}} & \multicolumn{2}{c}{$A(1^{1}\Sigma_{u}^{+})$} & \multicolumn{2}{c}{$B(1^{1}\Pi_{u})$} & \multicolumn{2}{c}{$2X(2^{1}\Sigma_{g}^{+})$} & \multicolumn{2}{c}{$C(1^{1}\Pi_{g})$} & \multicolumn{2}{c}{$b(1^{3}\Pi_{u})$} & \multicolumn{2}{c}{$c(1^{3}\Sigma_{g}^{+})$}\tabularnewline[2mm]
\hline 
\noalign{\vskip2mm}
\emph{ab initio }\cite{Musia2014} & 9356& & 2930& & 3289& & 1426& & \textbf{\textcolor{red}{12166}}& & 7080&\tabularnewline
empirical & 9353&1795(28) \cite{Gunton2013} & 2984&444(110) \cite{Huang2003} & \multicolumn{2}{c}{3318(66) \cite{He1991,Barakat1986} } & 1422&5(3) \cite{Miller1990} & \textbf{\textcolor{red}{12145}}&\textbf{\textcolor{red}{(200) }}\textcolor{black}{\cite{Engelke1983} } & 7092&4926(86) \cite{Semczuk2013}\tabularnewline
obs - calc & -3& & 54& & 30& & -3& & \textbf{\textcolor{red}{-21}}& & 12&\tabularnewline[2mm]
\hline 
\noalign{\vskip2mm}
\textbf{\textcolor{blue}{$2S+3S$}} & \multicolumn{4}{c}{$3X(3^{1}\Sigma_{g}^{+})$ (1st min.)} & \multicolumn{4}{c}{$2A(1^{1}\Sigma_{u}^{+})$ (1st min.)} & \multicolumn{4}{c}{$2A(1^{1}\Sigma_{u}^{+})$ (2nd min.)}\tabularnewline[2mm]
\hline 
\noalign{\vskip2mm}
\emph{ab initio }\cite{Musia2014} & \multicolumn{4}{c}{8290.} & \multicolumn{4}{c}{5608.} & \multicolumn{4}{c}{5389.}\tabularnewline
empirical & \multicolumn{4}{c}{8317. \cite{Bernheim1982,Bernheim1983} } & \multicolumn{4}{c}{5615. \cite{S2000,Kubkowska2007}} & \multicolumn{4}{c}{5321. \cite{S2000,Kubkowska2007}}\tabularnewline
obs - calc & \multicolumn{4}{c}{27.} & \multicolumn{4}{c}{7.} & \multicolumn{4}{c}{-68.}\tabularnewline[2mm]
\hline 
\noalign{\vskip2mm}
\textbf{\textcolor{blue}{$2S+3P$}} & \multicolumn{4}{c}{$4X(4^{1}\Sigma_{g}^{+})$ (1st min.)} & \multicolumn{2}{c}{$2B(2^{1}\Pi_{u})$} & \multicolumn{2}{c}{$2C(2^{1}\Pi_{g})$} & \multicolumn{2}{c}{$D(1{}^{1}\Delta_{g})$} & \multicolumn{2}{c}{$2d(2{}^{3}\Pi_{g})$}\tabularnewline[2mm]
\hline 
\noalign{\vskip2mm}
\emph{ab initio }\cite{Musia2014} & \multicolumn{4}{c}{8380.} & 6481& & 7773& & 9592& & 8505&\tabularnewline
empirical & \multicolumn{4}{c}{8349. \cite{Bernheim1981}} & 6455& \cite{Bernheim1981a} & 7773&7(3) \cite{Kubkowska2007} & 9579& \cite{Linton1993} & 8484& \cite{Bernheim1981a}\tabularnewline
obs - calc & \multicolumn{4}{c}{-31.} & -26& & 1& & -13& & -21&\tabularnewline[2mm]
\hline 
\end{tabular*}

\rule[0.1ex]{1\textwidth}{0.5pt}
\end{table*}

\subsection{Choice of model parameters}

Using the Hamiltonian of Eq. \ref{eq:hamiltonian}, we fit the parameters
of Eq. \ref{eq:VMLR} to the 1234 data, with the involved energy
levels of the upper states $2^{3}\Pi_{g}$ and $1^{3}\Delta_{g}$
treated as free parameters. All fits to the data were done using the
freely available program ${\tt DPotFit}\,2.0$ \cite{LeRoy2013}.
Starting parameters for the fits to Eq. \ref{eq:VMLR} were found
by fitting to an RKR potential using the freely available program
${\tt betaFit\,2.1}$ \cite{LeRoy2012}. The RKR potential was made
using the program ${\tt RKR1\,2.0}$ \cite{leroy2004} using the Dunham
coefficients found in Table IV of \cite{Linton1992}.

The quality of a fit was determined by the \textbf{d}imensionless
root-mean-square-\textbf{d}eviation ($\overline{dd}$) which scales
each deviation between an energy predicted by the model ($E_{{\rm calc}}$)
and the corresponding measurement ($E_{{\rm obs}})$, by the uncertainty
of the measurement ($u_{{\rm obs}}$), for all $N_{{\rm data}}$ measurements:

\begin{equation}
\overline{dd}\equiv\sqrt{\frac{1}{N_{{\rm data}}}\sum_{i=1}^{N_{{\rm data}}}\left(\frac{E_{{\rm calc}}(i)-E_{{\rm obs}}(i)}{u_{{\rm obs}}(i)}\right)^{2}}.
\end{equation}

In previous studies of the $A(1^{1}\Sigma_{u}^{+})$ state \cite{Gunton2013b}
and $c(1^{3}\Sigma_{g}^{+})$ state \cite{Semczuk2013}, it was determined
that there was no benefit in including long-range terms beyond $C_{8}$,
because the data for the high-lying rovibrational energies began to
deviate from the theoretical long-range potential energy curve at
distances shorter than the distance where $C_{9}$ began to give a
noticeable effect on the long-range function $u_{{\rm LR}}(r)$ (see
Fig. 6 of \cite{Semczuk2013} for example) . 

However, for the $b$-state, no data exists in the long-range region,
so it might make more sense to include more $C_{m}$ terms in $u_{{\rm LR}}(r)$
in order to anchor the potential somewhat appropriately in the $>$2500~cm$^{-1}$
gap at the top of the potential well where no data exists to guide
the potential. Nevertheless, we first followed the $A-$ and $c-$state
studies and only used up to $C_{8}$. We found an excellent fit with
$\overline{dd}=0.95$ with only $N_{\beta}=7$, $p=6,$ $q=2$. However,
the long-range behavior of this potential was in vast disagreement
with the long-range behavior expected by theory (see Fig. \ref{fig:leroySpace}).
This is because with $(p,q)=(6,2)$, the long-range form of the potential
described in Eq. \ref{eq:VMLR_longRange} does not ``turn on'' until
too high a value of $r$ (a larger value of $r$ is needed for $y_{p}^{r_{_{{\rm ref}}}}(r)$
and $y_{q}^{r_{_{{\rm ref}}}}(r)$ to become sufficiently close to
their limiting values of 1. 

We can often encourage the the long-range form to ``turn on'' earlier
by increasing $p$ and/or $q$, which often comes with the expense
of requiring a higher polynomial degree $N_{\beta}$ to recover the
same $\overline{dd}$. We explored models with $p\in\{6,7,8,9\}$
and $q\in\{2,3\}$, including the $C_{9}$ term in $u_{{\rm LR}}$
for $p\ge7$, $C_{10}$ for $p\ge8$, and $C_{11}$ for $p=9$. Ideally
we would always use as many $C_{m}$ constants as are known, but as
explained in \cite{LeRoy2009,Dattani2011}, the value of $p$ in Eqs.
\ref{eq:VMLR} and \ref{eq:betaPolynomial} need to satisfy $p>m_{{\rm last}}-m_{{\rm first}}$
where $m_{{\rm last}}$ and $m_{{\rm first}}$ represent respectively
the last and first $C_{m}$ terms included in $u_{{\rm LR}}(r)$.
We also note that $C_{12}$ for the Li$\left(2^{2}S\right)+\mbox{Li}\left(2^{2}P\right)$
asymptote is not available, as far as we know.

We found that if $p<9$ and/or $q<3$, the long-range behavior does
not turn on until about $r=32$\textcolor{black}{~$\mbox{\AA}$ (in
the very best cases), while the $m$-dependent Le~Roy radius \cite{Tsai1994}
calculated from the radial expectation values found in \cite{Zhang2007}
suggests that the long-range behavior should turn on before $r=10$~$\mbox{\AA}$.
With $(p,q)=(9,3)$, we found a fit with $N_{\beta}=11$ and $\overline{dd}=0.94$,
where the long-range behavior turns on at about }$r=20$\textcolor{black}{~$\mbox{\AA}$
(see Fig \ref{fig:leroySpace}). Increasing $q$ to 4 would likely
turn the long-range behavior on at closer to the $m$-dependent Le~Roy
radius, but no $(p,q)=(9,4)$ fits with $N_{\beta}\le11$ had a $\overline{dd}<1$
and we needed to push to $N_{\beta}\ge13$ in order to match the $\overline{dd}$
of the }best $(p,q,N_{\beta})=(9,3,11)$\textcolor{black}{{} fits. Using
such a high-degree polynomial, when the data only required $N_{\beta}=7$
for a good fit, can be dangerous in terms of the potential's extrapolation
in the regions neither constrained by data nor built-in $C_{m}$ constants.
In this respect, we also tried $(p,q)=(9,3)$ fits with $N_{\beta}=10$
for various $r_{{\rm ref}}$ values, but no such fit had a $\overline{dd}<1$. }

\subsection{Born-Oppenheimer breakdown (BOB) corrections \label{sub:Born-Oppenheimer-breakdown-(BOB)}}

With our best MLR model: MLR$_{p,q}^{r_{{\rm ref}}}\left(N_{\beta}\right)=\mbox{MLR\ensuremath{_{9,3}^{3.8}\left(11\right)}}$,
we attempted to add adiabatic ($\tilde{S}(r)$, from Eq. \ref{eq:firstTimeStildeIsIntroduced})
and non-adiabatic ($\tilde{R}(r)$, from Eq. \ref{eq:rotational_g_factor})
BOB corrections with the same model functions as used in previous
studies of Li$_{2}$ since these models were improved in 2009 \cite{LeRoy2009,Dattani2011a,Semczuk2013,Gunton2013}.
It was surprising that despite there being 599 $^{6}$Li$_{2}$ data
$\left(\mbox{with }v_{{\rm max}}=9,\, N_{{\rm max}}=46\right)$ and
696 $^{7}$Li$_{2}$ data $\left(\mbox{with }v_{{\rm max}}=27,\, N_{{\rm max}}=27\right)$,
adding BOB correction functions did not improve the fit. Even when
fitting to 3 adiabatic BOB parameters and 3 non-adiabatic BOB parameters,
the $\overline{dd}$ went down by less than 1\%. This is unexpected
when there is just as much data for each isotopologue, and there is
such a big difference in the highest $v$ and $N$ levels observed
for each isotopologue. 

Nevertheless, it seems that the isotopologue shifts due to the kinetic
energy term in the Hamiltonian, and due to the mass-dependent BOB
corrections incorporated from Eqs. \ref{eq:firstTermOfS(r)} and \ref{eq:P_LR},
are the only significant sources of energy difference between $^{6}$Li$_{2}$
and $^{7}$Li$_{2}$ for the $b$-state (within our data's precision).
This may also explain why the \emph{ab initio }potentials \cite{Musia2014}
calculated assuming an infinite molecular mass managed to predict
both the $^{6}$Li$_{2}$ and $^{7}$Li$_{2}$ energies so fabulously
(see discussion in Section \ref{sub:Vibrational-energy-spacings}
and Table IV), while the \emph{ab initio }BOB correction functions
for the not much lighter molecule BeH, were so crucial in matching
the ro-vibrational energies predicted from the \emph{ab initio }and
empirical potentials \cite{Dattani2015,Koput2011}. Therefore, the
final potential that we recommend, which is the same for both $^{6}$Li$_{2}$
and $^{7}$Li$_{2}$ in the $b$-state except for the small mass-dependent
contributions coming from the kinetic energy and the un-colored terms
in Eq. \ref{eq:DeltaV}, does not contain any empirically fitted $\tilde{S}(r)$
and $\tilde{R}(r)$ BOB correction functions.

\subsection{Sequential rounding and re-fitting (SRR)}

Observing the predicted values for $r_{e}$ yielded by 139 different
fits which had $\overline{dd}<0.957$ (within 1.5\% of the optimal
fit, which had $\overline{dd}=0.942$), we see that no fit predicted
an $r_{e}$ outside the range $(2.589\,825<r_{e}<2.589\,871)$~\textcolor{black}{$\mbox{\AA}$},
regardless of the values of $(p,q,r_{{\rm ref}},N_{\beta})$, though
the more extreme predictions of $r_{e}$ within this range corresponded
to fits with $N_{\beta}\le9$. Based on this observation, we recommend
the value $r_{e}=(2.589\,848\pm0.000\,023)$~\textcolor{black}{$\mbox{\AA}$},
which is the average of these upper and lower bounds, with the uncertainty
being the distance from the average to either bound.

We then re-fitted the potential to the data, but with $r_{e}$ fixed
at \textcolor{black}{2.589848~$\mbox{\AA}$, once with the ${\tt DPotFit}$
setting ${\tt IROUND=-1}$ and once with ${\tt IROUND=+1}$ in order
to implement the SRR procedure described in \cite{LeRoy1998} and
in the ${\tt DPotFit}$ manual \cite{LeRoy2013}. Neither of these
cases affected the 3-digit value $\overline{dd}=0.942$. The ${\tt IROUND=-1}$
fit ended up with 2 more total digits than when ${\tt IROUND=+1}$
was used, but had a lower $\overline{dd}$ in the 4th digit, and has
the more elegant feature that the number of digits in $\beta_{i}$
decreases monotonically with increasing $i$. Therefore, we recommend
the potential with ${\tt IROUND=-1}$, whose parameters are listed
in Table \ref{tab:parametersForPotential}. }

\subsection{Vibrational energy spacings of the recommended Li$_{2}\left(b,^{3}\Pi_{u}\right)$
potential, and comparison to best \emph{ab initio }potential\label{sub:Vibrational-energy-spacings}}

Very recently, a review paper on the 5e$^{-}$ systems BeH, BeD, and
BeT \cite{Dattani2015} revealed that the state of the art \emph{ab
initio }potentials \cite{Koput2011} (which used MR-ACPF/aug-cc-pCV7Z(i),
a further estimate of electron correlation effects beyond the approximations
of MR-ACPF, second-order DKH scalar relativistic corrections, and
mass-dependent BOB corrections), predicted vibrational energy spacings
with up to at most 1.8~cm$^{-1}$ discrepancy with the state of the
art empirical potential in the region for which vibrational energies
had been measured. The \emph{ab initio} potential also predicted the
existence of one more vibrational level than the empirical potential,
in the cases of BeH and BeD. This was all for the ground electronic
state $X\left(1{}^{2}\Sigma^{+}\right)$, so it is of interest to
see how well the most state-of-the-art \emph{ab initio }potential
for the 6e$^{-}$ Li$_{2}$ excited state $b\left(1{}^{3}\Pi_{u}\right)$
will be.

A Fock space MRCC method based on the (2,0) sector of the Fock space,
called FS-CCSD(2,0) \cite{Musial2012}, was recently implemented and
used to calculate potential energy curves for many excited states
of Li$_{2}$ \cite{Musia2014} with the ANO-RCC basis set \cite{Veryazov2004}.
While in principle possible, DKH and BOB corrections have not been
made in any Li$_{2}$ \emph{ab initio} calculations to date. However,
fortunately we found in Section \ref{sub:Born-Oppenheimer-breakdown-(BOB)}
that the addition of $\tilde{S}(r)$ or $\tilde{R}(r)$ functions
did not significantly improve the fit to the data, meaning that Born-Oppenheimer
breakdown beyond the effects included from Eqs. \ref{eq:firstTermOfS(r)}
and \ref{eq:P_LR} do not seem to have a big effect in this particular
state of Li$_{2}$, at least in the data region. Said another way,
the \emph{ab initio }Born-Oppenheimer potential is expected to give
good predictions of the energies of $^{6,6}$Li$_{2}$ and $^{7,7}$Li$_{2}$,
with mass-dependent differences accounted for only by the Hamiltonian's
kinetic energy operator, as was the case with the empirical MLR potential. 

\begin{table}
\protect\caption{{\scriptsize{}Parameters defining the recommended MLR}\textcolor{black}{\footnotesize{}$_{p,q}^{r_{{\rm ref}}}$}{\scriptsize{}
potential for the $b(1^{3}\Pi_{u})$-state of all isotopologues of
Li$_{2}$. The MLR model is defined in Eqs. \ref{eq:VMLR}-\ref{eq:damping (last MLR definition)},
and has damping parameters $s=-2$ and $\rho=0.5$. The appropriate
long-range constants are presented in Table \ref{tab:longRangeConstants}.
The fit's quality was $\overline{dd}=0.942$.\label{tab:parametersForPotential}}}

\rule{1\columnwidth}{0.5pt}

\begin{tabular*}{1\columnwidth}{@{\extracolsep{\fill}}>{\centering}p{0.25\columnwidth}c>{\raggedright}p{0.25\columnwidth}c>{\raggedright}p{0.4\columnwidth}}
\hline 
\noalign{\vskip2mm}
\multicolumn{5}{>{\centering}p{1\columnwidth}}{\textcolor{black}{MLR$_{9,3}^{3.80}$}}\tabularnewline[2mm]
\hline 
\noalign{\vskip2mm}
\multicolumn{5}{c}{~~~$\mathfrak{D}_{e}$\textcolor{black}{$\;\;12\,166$~~cm$^{-1}$}}\tabularnewline
\multicolumn{5}{c}{$r_{e}$\textcolor{black}{$\;\;2.589\,848(23)$~$\mbox{\AA}$}}\tabularnewline[2mm]
 & $\beta_{0}$ & \textcolor{black}{\,~$-0.022\,869\,54$} & $\beta_{6}$ & $-5.586\,9$\tabularnewline
 & $\beta_{1}$ & \textcolor{black}{\,~$-2.225\,706$} & $\beta_{7}$ & \textcolor{black}{\,}~~$7.787$\tabularnewline
 & $\beta_{2}$ & \textcolor{black}{\,~$-6.019\,341$} & $\beta_{8}$ & \textcolor{black}{\,\,}$25.55$\tabularnewline
 & $\beta_{3}$ & \textcolor{black}{$-10.672\,48$} & $\beta_{9}$ & \textcolor{black}{\,\,}$35.64$\tabularnewline
 & $\beta_{4}$ & $-13.746\,79$ & $\beta_{10}$ & \textcolor{black}{\,\,}$26.5$\tabularnewline
 & $\beta_{5}$ & $-12.632\,2$ & $\beta_{11}$ & \textcolor{black}{\,}~~$8.4$\tabularnewline[2mm]
\hline 
\end{tabular*}

\rule{1\columnwidth}{0.5pt}
\end{table}

Using the \emph{ab initio }Born-Oppenheimer potential provided to
us by the authors of \cite{Musia2014}, and the MLR potential described
by Table \ref{tab:parametersForPotential}, we used ${\tt LEVEL}$
to calculate the vibrational energies of both the $^{6,6}$Li$_{2}$
and $^{7,7}$Li$_{2}$ isotopologues.  We found that the highest
levels had outer classical turning points of several thousand Angstroms,
and therefore we found it useful to use the recently developed mapping
which allows the radial mesh to extend to $r=\infty$ when numerically
solving the Schroedinger equation \cite{Meshkov2008,Meshkov2011},
which is also implemented in ${\tt LEVEL}$. With this method we were
able to find up to $v=91$ for $^{6,6}$Li$_{2}$ and $v=98$ for
$^{7,7}$Li$_{2}$, however, when we calculated the scattering wavefunction,
the number of nodes indicated that the highest bound vibrational levels
should be $v=92$ and $v=100$ respectively. Impressively, these results
were identical whether we used the \emph{ab initio} potential, or
the MLR potential. 

We used Le~Roy-Bernstein theory to predict these missing levels for
each isotopologue: For a $\nicefrac{C_{3}}{r^{3}}$ potential, the
powers $E_{v}^{\left(\nicefrac{1}{6}\right)}$ of the binding energies
should be linear in $v$ \cite{Leroy1970}. We used the slope calculated
from $v=90$ and $91$ for $^{6,6}$Li$_{2}$, and the slope calculated
from $v=97$ and $98$, for predicting the energies of $v=92$ and
$v=99$ levels respectively. We then used the last two points again
to calculate a new slope for predicting the energies of $v=100$.
Interestingly, both the \emph{ab initio} potential, and the MLR potential
predict the existence of a $^{6,6}$Li$_{2}$ level bound by $<8\times10^{-8}$~cm$^{-1}$
($<3$~kHz) and a $^{7,7}$Li$_{2}$ level bound by $<2\times10^{-10}\,$cm$^{-1}$
($<6$~Hz). Using $C_{3}^{^{6}{\rm Li}}/r^{3}$ and $C_{3}^{^{7}{\rm Li}}/r^{3}$
we get that the outer classical turning points for the least bound
levels of each isotopologue are predicted to be at least 13~000~\textcolor{black}{$\mbox{\AA}$}
and 120~000~\textcolor{black}{$\mbox{\AA}$} respectively.

These vibrational energies were then used to calculate the zero point
energies (ZPEs) and vibrational energy spacings $\omega_{i}$, which
are presented in the table below, along with the discrepancy between
the \emph{ab initio} and empirical potentials. We have compared the
vibrational energies (since these are important for photoassociation
experiments) and the vibrational spacings (since these are important
for experiments involving energy transitions). For both $^{6,6}$Li$_{2}$
and $^{7,7}$Li$_{2}$, the discrepancy for the vibrational energies
is less than 12~cm$^{-1}$. The agreement for the vibrational spacings
is much better than for the case of BeH discussed in the beginning
of this subsection. The largest discrepancy for a $^{6,6}$Li$_{2}$
vibrational spacing is $<0.8\,$cm$^{-1}$ and for $^{7,7}$Li$_{2}$~cm$^{-1}$
is $<0.6\,$cm$^{-1}$.

\section{Conclusion}

The motivation for this work was to build a potential that could predict
high-accuracy vibrational energies for $^{6,6}$Li$_{2}(b)$ in the
accessible energy range of the recently built high-precision experimental
setup which has so far been very successful for photoassociation spectroscopy
of $c\left(1^{3}\Sigma_{g}^{+}\right)$ \cite{Semczuk2013} and $A\left(1^{1}\Sigma_{u}^{+}\right)$-states
\cite{Gunton2013}. A similar photoassociation apparatus has recently
also been setup by Kai Dieckmann's group to measure energy levels
of ultra-cold $^{6,6}$Li$_{2}$ electronic states dissociating to
the $2S+3P$ asymptote \cite{Sebastian2014}. The best \emph{ab initio}
vs empirical potential comparison for Li$_{2}$ in the literature
\cite{Halls2001}, predicted vibrational levels with a discrepancy
of up to 2.04~cm$^{-1}$ for the $a\left(1^{3}\Sigma_{u}^{+}\right)$,
which would have simply not been good enough for finding the levels
in this type of experiment. The spectroscopic features in this type
of experiment are typically around 0.000~2~cm$^{-1}$; and covering
2~cm$^{-1}$ with one-minute measurements and a 0.000~2~cm$^{-1}$
step size would take about 7 days. 

However, the empirical MLR potential of \cite{Dattani2011} for the
$c$-state predicted energies were accurate enough to cut the experiment's
duration to under 2 days, since the first level in the laser's range
turned out to be predicted correctly to within 0.525~cm$^{-1}$,
despite this energy being right in the middle of a 5000~cm$^{-1}$
gap in available experimental data to guide the empirical potential.
In our table comparing the \emph{ab initio} and empirical MLR energies
for the $b$-state we see that the vibrational energies predicted
by the \emph{ab initio} potential are sometimes in $>10$~cm$^{-1}$
disagreement with the empirical values in the region where the energies
have in fact been measured. However, the \emph{ab initio }seems to
predict all vibrational energy spacings correctly within less than
1~cm$^{-1}$ which is much better than the result in the current
best ground state 5e$^{-}$ BeH study \cite{Dattani2015,Koput2011}. 

The reason we are interested in measuring more levels of the $b$-state
with high-precision, is because it is surprising that the best \emph{ab
initio} calculation of the first Li$(2^{2}S)-\mbox{Li\ensuremath{(2^{2}P)}}$
interaction term ($C_{3}$) is still in vast disagreement with the
empirically fitted values from the studies of the $A$-state \cite{LeRoy2009,Semczuk2013}
and $c$-state \cite{Dattani2011,Gunton2013}, despite lithium only
having 3e$^{-}$, and this $C_{3}$ value  having significance for
atomic clocks \cite{Tang2011}. Lithium is also expected to play a
major role in polarizability metrology, since polarizability ratios
can be measured much more precisely than individual polarizabilities
\cite{Cronin2009} and Li is the preferred choice for the standard
in the denominator of such a ratio \cite{Mitroy2010}. But this discrepancy
in $C_{3}$ limits the accuracy of a potential Li-based standard for
polarizabilities \cite{Tang2011}. Consolingly, in this study we have
found that the $b$-state is predicted to have levels bound by $<8\times10^{-8}$~cm$^{-1}$
($<3$~kHz) which would imply an outer classical turning point of
$>13\,000\,$\AA , which is larger than any case in our awareness.
Since the less bound the level measured, the more precisely $C_{3}$
can be determined from a fit, these extremely weakly bound energies
are promising for resolving the discrepancy. While the technology
to measure these extremely weakly bound energies may still be years
away, many of the very high vibrational levels predicted in our analysis
\emph{are} indeed accessible with today's photoassociation technology. 

The least bound levels for the $A$-state which have been measured
have binding energies of $E_{v=83}\approx6$~cm$^{-1}$, $E_{v=88}\approx0.7$~cm$^{-1}$
and $E_{v=97}\approx0.4$~cm$^{-1}$ for $^{6.7}$Li$_{2}$, $^{6,6}$Li$_{2}$
and $^{7,7}$Li$_{2}$ respectively, and the least-squares fit to
the data gave a $C_{3}$ value with a 95\% confidence limit uncertainty
of about $\pm$8~cm$^{-1}/\text{\AA}^{3}$ \cite{LeRoy2009}, which
is currently the most precise experimentally determined oscillator
strength for any system, by an order of magnitude \cite{Tang2011}.
The \emph{ab initio} and empirical MLR potentials for the $b$-state
compared in this work, both give predictions that are in great agreement
for energy levels that are several orders of magnitude less deeply
bound than the least deeply bound $A$-state measurements, making
it therefore possible to obtain an empirical $C_{3}$ value far more
precise than in \cite{LeRoy2009}. Hopefully, this would resolve the
age-old discrepancy between experiment and theory for this $C_{3}$
value, which was first measured experimentally by Loomis and Nusbaum
in 1931 \cite{Loomis1931}. 

Empirical potentials have recently been built for the $b$- and $A$-states
of: Rb$_{2}$ in 2009 \cite{Salami2007} and again in 2013\cite{Drozdova2013},
NaCs in 2009 \cite{Zaharova2009}, KCs in 2010 \cite{Kruzins2010},
RbCs in 2010 \cite{Docenko2010}, Cs$_{2}$ in 2011 \cite{Bai2011}
and NaK in 2015 \cite{Harker2015}, however, this is to our knowledge,
the frst empirical potential built for the $b$-state of Li$_{2}$.

\begin{widetext}

\textcolor{black}{\scriptsize{}Table IV. Comparison of the binding
energies, denoted $G(v_{i})$; zero-point energies (ZPE); and vibrational
energy spacings, denoted $\omega_{i}\equiv G(v_{i+1})-G(v_{i})$;
for $^{6,6}$Li$_{2}$ and $^{7,7}$Li$_{2}$. The last column is
the difference between the two columns directly prior. Discrepancies
of $\ge$ 0.5~cm$^{-1}$ are marked by one star (two stars if it
was for vibrational level within the data range). Lines with blue
font are for unobserved levels, lines with bold green font are for
unobserved levels which are accessible by the UBC lab.All energies
were calculated by the program $\texttt{LEVEL 8.2}$ with atomic masses
and $\texttt{NUSE=0}$, $\texttt{IR2=1}$, $\texttt{ILR=3}$, $\texttt{NCN=6}$,
and $\texttt{CNN=}C_{3}$. All numbers were converged with respect
to the radial mesh parameters, to at least to the number of digits
shown. \label{tab:ComparisonOfVibrationalEnergies-1}}{\scriptsize \par}

\begin{longtable}{crrcccrr>{\centering}p{19mm}>{\raggedright}p{8mm}}
\hline 
\noalign{\vskip2mm}
\multirow{2}{*}{\textcolor{black}{\small{}~$v$}} & \textcolor{black}{\small{}2015 Empirical } & \textcolor{black}{\small{}2014 ~~~~~~~} & \multicolumn{2}{c}{} & \multirow{2}{*}{} & \textcolor{black}{\small{}2015 Empirical \negmedspace{}} & \textcolor{black}{\small{}2014 ~~~~~~} & \multicolumn{2}{c}{}\tabularnewline
 & \textcolor{black}{\small{}{[}Present work{]}~}\textcolor{blue}{\small{}\,} & \textcolor{black}{\emph{\small{}ab initio~~~~~}} & \multicolumn{2}{c}{{\small{}$E_{v}(2015)-E_{v}(2014)$}} &  & \textcolor{black}{\small{}{[}Present work{]}}\textcolor{blue}{\small{}\,} & \textcolor{black}{\emph{\small{}ab initio~~~~}} & \multicolumn{2}{c}{\textcolor{black}{\small{}$\omega(2015)-\omega(2014)$}}\tabularnewline[2mm]
\hline 
\hline 
\noalign{\vskip2mm}
\multicolumn{3}{c}{} & \multicolumn{7}{l}{~~~~~~~~~~~~~~~~~~~~~\textcolor{black}{$^{6,6}$Li$_{2}$}}\tabularnewline[2mm]
\hline 
\noalign{\vskip2mm}
\textbf{\small{}$\mathfrak{D}_{e}$} & {\small{}-12166~~~~~~~~~~~~~} & {\small{}-12166~~~~~~~~~~~~~} & {\small{}-} &  & {\small{}-} & {\small{}-~~~~~~~} & {\small{}-~~~~~~~~~} & {\small{}-} & \tabularnewline[-0.5mm]
\noalign{\vskip-0.15mm}
\textbf{\textcolor{black}{\small{}$0$}} & {\small{}-11977.92296303} & {\small{}-11978.33396628} & {\small{}0.00} &  & {\small{}ZPE} & {\small{}188.07703700} & {\small{}187.66603}\textcolor{blue}{\small{}\,}{\small{}370} & {\small{}0.41} & \tabularnewline[-0.5mm]
\noalign{\vskip-0.15mm}
\textbf{\textcolor{black}{\small{}$1$}} & {\small{}-11608.64518318} & {\small{}-11609.63836960} & {\small{}0.41} &  & \textbf{\textcolor{black}{\small{}$\omega_{0}$}} & {\small{}369.27777985} & {\small{}368.69559}\textcolor{blue}{\small{}\,}{\small{}668} & {\small{}0.58} & {\scriptsize{}{*}{*}}\tabularnewline
\hline 
\end{longtable}

\end{widetext}

\clearpage{}

\bibliographystyle{IEEEtran}

\end{document}